\documentclass[12pt]{article}
\usepackage{graphicx}
\usepackage{amsmath}
\usepackage{amssymb}
\usepackage[numbers,compress]{natbib}
\usepackage{amsmath}
\begin{document}

\title{Longitudinal Doppler effect in de Sitter expanding universe}

\author{Ion I. Cot\u aescu\\ {\small \it  West 
                 University of Timi\c soara,}\\
   {\small \it V. P\^ arvan Ave. 4, RO-1900 Timi\c soara, Romania}}

\maketitle
\begin{abstract}
A new redshift formula is obtained considering  the longitudinal Doppler effect in the de Sitter expanding universe where the relative geodesic motion is governed by  the Lorenzian isometries  of our new de Sitter relativity [I. I. Cot\u aescu, {Eur. Phys. J. C} {\bf 77} (2017) 485]. This formula combines in a non-trivial manner  the well-known cosmological contribution given by Lama\^ itre's law with that of the relative  motion of the source with respect to a fixed observer as in de Sitter relativity instead of special relativity.  Other related quantities  as the dispersion relation, the propagation time of the photon and the real distance between source and observer at the moment of observation  are discussed pointing out their specific features in the de Sitter relativity. 

Pacs: 04.02.Cv and  04.02
\end{abstract}

\section{Introduction}

In general relativity the Doppler effect can be seen as being produced by the expansion of the universe or by the relative motion as in special relativity. The cosmological effect complies with the general redshift law  \cite{SW,H0,H} while the kinetic one is treated separately  by using the methods of special relativity \cite{LL}. Thus possible interferences of these two effects cannot be pointed out without resorting to a more general theory of relativistic effects in the presence of gravity as our de Sitter relativity  we proposed recently \cite{CdSR1,CdSR2}.  This gives us the opportunity of analysing the Doppler effect globally considering simultaneously the cosmological and kinetic contributions in the de Sitter expanding universe.

In de Sitter relativity the role of inertial frames is played by any set of local charts related trough isometries. In what follows we focus mainly on the comoving charts with conformal coordinates where the Maxwell equations have similar solutions as in Minkowski spacetime allowing us to define correctly the photon energy and momentum \cite{Max}. For studying the relative motion of these frames we exploit the Lorentzian isometries defined recently \cite{CdSR1} and the corresponding transformation rules of the conserved quantities. Thus we may deduce the observed energies in different frames  deriving the Doppler effect in de Sitter relativity. Note that in the de Sitter manifold the energy and momentum transform under isometries as the components of a five-dimensional skew-symmetric tensor in association with the angular momentum and a new specific conserved vector we called adjoint momentum \cite{CGRG,CdSR1}.  For this reason the formalism is different from the usual one of special relativity but the philosophy of the relative motion remains the same.

Since here we make the first step to this approach we restrict ourselves to the longitudinal Doppler effect in which the source carried by the mobile frame is translated with the distance $d$ from its origin only along the direction of the relative velocity. Our goal is to analyse how a photon emitted at the initial time by this source is observed by a fixed observer when we know that at that time the origin of the mobile frame is passing through the origin of the fixed one with the relative velocity ${\bf V}$.  Thus we formulate a problem of relative motion in the presence of the de Sitter expansion able to reveal the interference between the cosmological and kinetic contributions to the Doppler effect.  Applying this method we obtain a new formula of this effect such that in the particular case of ${\bf V}=0$ we recover the cosmological effect given by Lema\^ itre's law \cite{L1,L2} while for $d=0$ we find just the well-known formula of the Doppler effect in special relativity.  Moreover, we derive the related significant quantities namely the dispersion relation, the propagation time of the photon and the real distance between source and observer at the moment of observation.

The paper is organized as follows. In the second section we briefly present the de Sitter relativity in conformal and respectively de Sitter-Painlev\' e local charts showing how the coordinates and conserved quantities transform under isometries. The next section is devoted to the longitudinal Doppler effect for which we give the concrete form of the Lorentzian isometries deriving the corresponding transformation rules of the conserved quantities leading to the final formula relating the emitted and observed photon frequencies. Moreover, we analyse the mentioned related quantities pointing out their specific features resulted from the de Sitter relativity \cite{CdSR1} which uses Lorentzian isometries instead of the usual boosts of special relativity.  Finally we present few concluding remarks.

\section{de Sitter relativity}

Let us start with the de Sitter spacetime $(M,g)$ defined as the hyperboloid of radius $1/\omega$  in the five-dimensional flat spacetime $(M^5,\eta^5)$ of coordinates $z^A$  (labeled by the indices $A,\,B,...= 0,1,2,3,4$) having the metric $\eta^5={\rm diag}(1,-1,-1,-1,-1)$. The local charts $\{x\}$  of coordinates $x^{\mu}$ ($\alpha,\mu,\nu,...=0,1,2,3$) can be introduced on $(M,g)$ giving the set of functions $z^A(x)$ which solve the hyperboloid equation,
\begin{equation}\label{hip}
\eta^5_{AB}z^A(x) z^B(x)=-\frac{1}{\omega^2}\,.
\end{equation}
where  $\omega$ denotes the Hubble de Sitter constant since in our notations   \cite{CGRG}.

The de Sitter isometry group is just the gauge group $SO(1,4)$ of the embedding manifold $(M^5,\eta^5)$  that leave  invariant its metric and implicitly Eq. (\ref{hip}). Therefore, given a system of coordinates defined by the functions $z=z(x)$, each transformation ${\frak g}\in SO(1,4)$ defines the isometry $x\to x'=\phi_{\frak g}(x)$ derived from the system of equations
\begin{equation}\label{zz}
z[\phi_{\frak g}(x)]={\frak g}z(x)
\end{equation}
that holds for any type of coordinates which means that  these isometries are defined globally. The sets of local charts related through these isometries play the role of the systems of  inertial frames similar to those of special relativity. 

In what follows we consider the comoving charts with two sets of local coordinates,  the {\em conformal} pseudo-Euclidean ones, $\{t_c,{\bf x}_c\}$, and the 'physical' de Sitter-Painlev\' e coordinates, $\{t,{\bf x}\}$. The conformal time $t_c$ and Cartesian spaces coordinates $x_c^i$ ($i,j,k,...=1,2,3$) are defined by the functions 
\begin{eqnarray}
z^0(x_c)&=&-\frac{1}{2\omega^2 t_c}\left[1-\omega^2({t_c}^2 - {\bf x}_c^2)\right]\,,
\nonumber\\
z^i(x_c)&=&-\frac{1}{\omega t}x_c^i \,, \label{Zx}\\
z^4(x_c)&=&-\frac{1}{2\omega^2 t_c}\left[1+\omega^2({t_c}^2 - {\bf x}_c^2)\right]\,,
\nonumber
\end{eqnarray}
written with the vector notation, ${\bf x}=(x^1,x^2,x^3)\in {\Bbb R}^3\subset M^5$. These charts  cover the expanding part of $M$ for $t_c \in (-\infty,0)$
and ${\bf x}_c\in {\Bbb R}^3$ while the collapsing part is covered by
similar charts with $t_c >0$. In both these cases we have the same conformal flat line element,
\begin{equation}\label{mconf}
ds^{2}=\eta^5_{AB}dz^A(x_c)dz^B(x_c)=\frac{1}{\omega^2 {t_c}^2}\left({dt_c}^{2}-d{\bf x}_c\cdot d{\bf x}_c\right)\,.
\end{equation}
In what follows we restrict ourselves to the expanding portion which is a plausible  model of our expanding universe. 

The  de Sitter-Painlev\' e coordinates $\{t, {\bf x}\}$ on the expanding portion  can be introduced  directly by substituting
\begin{equation}\label{EdS}
t_c=-\frac{1}{\omega}e^{-\omega t}\,, \quad {\bf x}_c={\bf x}e^{-\omega t}\,,
\end{equation}
where $t\in(-\infty, \infty)$ is the {proper} or cosmic time while $x^i$ are the 'physical' Cartesian space coordinates. Then the line element reads
\begin{equation}
ds^2=(1-\omega^2 {{\bf x}}^2)dt^2+2\omega {\bf x}\cdot d{\bf x}\,dt -d{\bf x}\cdot d{\bf x}\,. 
\end{equation}
Notice that this chart is useful in applications since in the flat limit (when $\omega \to 0$) its coordinates become just the Cartesian ones of the Minkowski spacetime.  

In the charts with combined coordinates $\{t,{\bf x}_c\}$ the metric takes the  Friedman-Lema\^ itre-Robertson-Walker (FLRW) form 
\begin{equation}
ds^2=dt^2-a(t)^2\,d{\bf x}_c\cdot d{\bf x}_c\,, \quad a(t)=e^{\omega t}\,, 
\end{equation}
where $a(t)$ is the scale factor of the expanding portion which can be rewritten in the conformal chart,  
\begin{equation}\label{scale}
a(t_c)\equiv a[t(t_c)]=-\frac {1}{\omega t_c}\,,
\end{equation}
as a function defined for $t_c<0$.

The classical conserved quantities under de Sitter isometries  can be calculated with the help of  the  Killing vectors  $k_{(AB)}$ of the de Sitter manifold $(M,g)$ \cite{CGRG}. According to the general definition of the Killing vectors in the pseudo-Euclidean spacetime $(M^5,\eta^5)$, we may consider the following identity
\begin{equation}
K^{(AB)}_Cdz^C=z^Adz^B-z^Bdz^A=k^{(AB)}_{\mu}dx^{\mu}\,,
\end{equation}
giving the covariant components  of the Killing vectors in an arbitrary chart $\{x\}$ of  $(M,g)$ as
\begin{equation}\label{KIL}
k_{(AB)\,\mu}=\eta^5_{AC}\eta^5_{BD}k^{(CD)}_{\mu}= z_A\partial_{\mu}z_B-z_B\partial_{\mu}z_A\,, 
\end{equation}
where $z_A=\eta_{AB}z^B$. The principal conserved quantities along the timelike geodesics  have the general form  ${\cal K}_{(AB)}(x,{\bf P})=\omega k_{(AB)\,\mu}m u^{\mu}$ where $u^{\mu}=\frac{dx^{\mu}(s)}{ds}$ are the components of the covariant four-velocity that satisfy  $u^2=g_{\mu\nu}u^{\mu}u^{\nu}=1$. 

We have shown that in a conformal chart $\{t_c,{\bf x}_c\}$ the geodesic equation  of a particle of mass $m$ passing through the space point ${\bf x}_{c0}$ at time ${t_{c0}}$  is completely determined by the initial condition  ${\bf x}_c({t_{c0}}) ={\bf x}_{c0}$  and the conserved momentum ${\bf P}$ as \cite{CGRG,CdSG},
\begin{equation}
{x_c}^i(t_c)={x_c}_0^i+\frac{P^i}{\omega {P}^
2} \left(\sqrt{m^2+{P}^{2}\omega^2 {t_{c0}}^2}-\sqrt{ m^2+{P}^2
\omega^2 t_c^2}\, \right)\,,\label{geodE}
\end{equation}
where we denote $P=|{\bf P}|$. Moreover, the  other conserved quantities  in an arbitrary point $(t_c,{\bf x}_c(t_c))$ of the geodesics  depend only on this point and the momentum ${\bf P}$.  These are the energy $E$, angular momentum ${\bf L}$ and a specific vector ${\bf Q}$ we called the adjoint momentum. In the chart $\{t_c,{\bf x}_c\}$ these quantities have the form \cite{CdSR1,CdSG}
\begin{eqnarray}
E&=&\omega\, {\bf x}_c(t_c)\cdot {\bf P}+\sqrt{ m^2+{P}^{2}\omega^2 t_c^2}\,,\label{Ene}\\
L_i&=&\varepsilon_{ijk} x^j_c(t_c) P^k\,,\label{L}\\
Q^i&=&2\omega x_c^i(t_c)E+\omega^2P^i[t_c^2-{\bf x}_c(t_c)^2]\,.\label{Q}
\end{eqnarray}
satisfying the  obvious identity
\begin{equation}\label{disp}
E^2-\omega^2 {{\bf L}}^2-{\bf P}\cdot {\bf Q}=m^2
\end{equation}
corresponding to the first Casimir invariant of the $so(1,4)$ algebra \cite{CGRG}. In the flat limit, $\omega\to 0$  and $-\omega t_c\to 1$, we have ${\bf Q} \to {\bf P}$ such that this identity  becomes just  the usual mass-shell condition $E^2-{\bf P}^2=m^2$ of special relativity.   

The conserved quantities $E$, ${\bf P}$ and the new ones,
\begin{equation}\label{KR}
{\bf K}=-\frac{1}{2\omega}\left({\bf P}-{\bf Q}\right)\,, \quad {\bf R}=-\frac{1}{2\omega}\left({\bf P}+{\bf Q}\right)\,,
\end{equation}
form a   skew-symmetric tensor on $M^5$,  
 \begin{equation}
{\cal K}(x,{\bf P})=
\left(
\begin{array}{ccccc}
0&\omega K_1&\omega K_2&\omega K_3&E\\
-\omega K_1&0&\omega L_3&-\omega L_2&\omega R_1\\
-\omega K_2&-\omega L_3&0&\omega L_1&\omega R_2\\
-\omega K_3&\omega L_2&-\omega L_1&0&\omega R_3\\
-E&\omega R_1&-\omega R_2&-\omega R_3&0
\end{array}\right)\,,\label{KK}
\end{equation}
whose elements  transform under an isometry $x\to x_c'=\phi_{\frak g}(x_c)$ defined by Eq. (\ref{zz}) as  
\begin{equation}\label{KAB}
{\cal K}_{(AB)}(x_c',{{\bf P}}')={\frak g}_{A\,\cdot}^{\cdot\,C}\,{\frak g}_{B\,\cdot}^{\cdot\,D}\,{\cal K}_{(CD)}(x_c,{\bf P})\,,
\end{equation}
for all ${\frak g}\in SO(1,4)$. Here  ${\frak g}_{A\,\cdot}^{\cdot\,B}=\eta^5_{AC}\,{\frak g}^{C\,\cdot}_{\cdot \,D}\, \eta^{5\,BD}$ are the matrix elements of the adjoint matrix $\overline{\frak g}=\eta^5\,{\frak g}\,\eta^5$. Thus,  Eq. (\ref{KAB}) can be written as  ${\cal K}(x',{{\bf P}}')=\overline{\frak g}\,{\cal K}(x,{\bf P})\,\overline{\frak g}^T$ or simpler, ${\cal K}'=\overline{\frak g}\,{\cal K}\,\overline{\frak g}^T$. 

Concluding, we can say that the de Sitter isometries are generated globally by the $SO(1,4)$ transformations which determine simultaneously  the transformations of the coordinates and of the conserved quantities.  We have thus a specific relativity on the de Sitter spacetime allowing us to study different relativistic effects in the presence of the de Sitter gravity. 

\section{Longitudinal Doppler effect}

Let us consider two observers, the first one staying at rest in the origin $O$ of the fixed frame $\{t_c,{\bf x}_c\}$ and the second one staying in the origin $O'$ of a mobile frame $\{t'_c,{\bf x}'_c\}$ moving along the $x^1_c$ axis. We adopt the synchronization condition assuming that $O'$ passes through $O$ with the velocity ${\bf V}=(V,0,0)$ at the initial moment
\begin{equation}\label{ini}
t_{c0}=t_{c0}'=-\frac{1}{\omega} ~~~\to~~~ t_0=t_0'=0\,.
\end{equation}
Then, assuming that the observer $O$ measures the parameters $({t_{c}},{\bf x}_{c},{\bf P})$ while $O'$ observes other parameters, $(t'_{c},{\bf x}'_{c},{\bf P}')$, of the same particle, we may apply the general results of our de Sitter relativity.

\subsection{Lorenzian isometry}

In what follows we study the relative motion by  using the {\em Lorenzian isometries} defined in Ref. \cite{CdSR1} instead of the usual Lorentz transformations of special relativity. In our particular case these de Sitter isometries, are generated by the particular Lorentz transformation along the $z^1$ axis (which is parallel with the $x_c^1$ axis)  of the form  
\begin{equation}
{\frak g}({\bf V})=\left(
\begin{array}{ccccc}
\frac{1}{\sqrt{1-V^2}}&\frac{V}{\sqrt{1-V^2}}&0&0&0\\
\frac{V}{\sqrt{1-V^2}}&\frac{1}{\sqrt{1-V^2}}&0&0&0\\
0&0&1&0&0\\
0&0&0&1&0\\
0&0&0&0&1
\end{array}\right) \,.
\end{equation}
Note that this is a transformation of the $SO(1,4)$ group acting  in $M^5$ that may not be confused with an usual Lorentz boost of special relativity. With their help we can derive the isometry $x_c=\phi_{{\frak g}({\bf V})}(x_c')$ by solving the system (\ref{zz}) for ${\frak g}={\frak g}({\bf V})$ and the functions (\ref{Zx}). We obtain  \cite{CdSR1},
\begin{eqnarray}
t_c(t_c',{\bf x}'_c)&=&\frac{t'_c}{\Delta_c'}\,,\label{Eq1}\\
{{x}^1_c}(t_c',{\bf x}^{\prime}_c)&=&\frac{1}{\Delta_c'}\left\{\gamma {x}^{\prime\, 1}_c+\frac{\gamma V}{2\omega}\left[1-\omega^2\left((t_c')^2-{{\bf x}'_c}\cdot{{\bf x}'_c}\right)\right]\right\}\,,\label{Eq2}\\
{{x}^2_c}(t_c',{\bf x}^{\prime}_c)&=&\frac{ \gamma {x}^{\prime\,2}_c}{\Delta_c'}\,,\\
{{x}^3_c}(t_c',{\bf x}^{\prime}_c)&=&\frac{\gamma {x}^{\prime\,3}_c}{\Delta_c'} \,,\label{Eq4}
\end{eqnarray}
where $\gamma=(1-V^2)^{-\frac{1}{2}}$ and
\begin{equation}
\Delta_c'=1+\omega\gamma\, {\bf x}'_c\cdot{\bf V}+\frac{\gamma-1}{2}\left[1-\omega^2\left((t_c')^2-{{\bf x}'_c}\cdot{{\bf x}'_c}\right)\right]\,.
\end{equation} 
The conserved quantities put in the form (\ref{KK}) with the help of Eqs. (\ref{KR}) transform under this isometries  as  \cite{CdSR1}
\begin{equation}\label{Kg}
{\cal K}(t_c,{\bf x}_c,{\bf P})=\overline{\frak g}({\bf V})\,\,{\cal K}(t_c',{\bf x}'_c,{\bf P}')\overline{\frak g}({\bf V})^T\,,
\end{equation}
as it results from Eq. (\ref{KAB}).  Thus we have all the transformation rules relating the coordinates and the conserved quantities observed by $O'$ and $O$ whose frames are related through a Lorentzian isometry of parameter ${\bf V}$. Note that in de Sitter relativity this parameter represent the relative velocity only at the initial time (\ref{ini}) since in this geometry the relative velocities of the geodesic motions depend on time.
   
In special relativity an electromagnetic source carried by a mobile frame produces the same Doppler effect regardless its fixed position with respect of this frame since in the Minkowski spacetime the energy is independent on translations.  In contrast, in de Sitter relativity the energy depends on position as in Eq. (\ref{Ene}) which means that the Doppler effect will depend on the position of the source with respect to the mobile frame. 

In order to avoid complicate calculations here we restrict ourselves to the particular case of the longitudinal Doppler effect when the source is translated only along the direction of the velocity of the mobile frame $\{t'_c,{\bf x}_c'\}$, staying at rest in the space point  $(d,0,0)$ of this frame. The observer $O'$ and implicitly $O$ will receive signals from this source only if this remains inside the null cone in $t_{c0}=t'_{c0}=-\omega^{-1}$ such that the condition $\omega d<1$ is mandatory \cite{CdSG}.

We must specify that in the associated frame with de Sitter-Painlev\' e coordinates  $\{t',{\bf x}'\}$ defined by Eq. (\ref{EdS}) the source is moving because of the manifold expansion, having the coordinates $(t',d(t'),0,0)$ where  $d(t')=d e^{\omega t'}$. Bering in mind that we fixed the initial moment when $O=O'$ as in Eq. (\ref{ini}) we understand that $d$ is just the physical distance of the source moving with respect to the mobile frame with the velocity 
\begin{equation}
v=\left.\frac{d d(t')}{dt'}\right|_{t'=0}=\omega d\,,
\end{equation}
at the initial time $t_0=t_0'=0$. Thus we recover the well-known velocity-distance law  (in the mobile frame) that occurs naturally in the de Sitter expanding universe as in any other FLRW spacetime \cite{H}.

\subsection{Emitted photon}
  
In the conformal charts the electromagnetic potential has the same form as in the Minkowski spacetime since the Maxwell equations are invariant under conformal transformations \cite{Max}. This means that a regressive plane electromagnetic wave, that has to be observed successively by $O'$ and $O$, may have the momentum ${\bf k}=(-k,0,0)$ and energy $k$ being proportional with
\begin{equation}
{A}_i\propto \varepsilon_i e^{-i k x^1_c-ik t_c}
\end{equation}
where $\varepsilon_i$ are the components of an arbitrary polarisation vector. Assuming that a photon of this type is emitted at the initial time (\ref{ini}) we find the form of its geodesic equation in the frame $O'$  \cite{CdSG}
\begin{equation}
x_{c\,ph}^{\prime\,1}(t_c')=d-\frac{1+\omega t_c'}{\omega}\,,\quad x_{c\,ph}^{\prime\,2}(t_c')=x_{c\,ph}^{\prime\,3}(t_c')=0\,.
\end{equation}
Then by using the transformation equations (\ref{Eq1}) - (\ref{Eq4}) we can derive the parametric equations of the photon geodesic observed by $O$, 
\begin{eqnarray}
t_{ph}(t_c')&=&t\left[t_c',{\bf x}'_{c\,ph}(t'_c)\right]\,,\label{G1}\\
{\bf x}_{ph}(t_c')&=&\bf{x}\left[t_c',{\bf x}'_{c\,ph}(t'_c)\right]\,,\label{G2}
\end{eqnarray}
depending on the parameter $t'_c$.

The conserved quantities related to the photon trajectory are observed differently by the mobile and fixed observers.  According to Eq. (\ref{Ene}),  the observer $O'$ measures the photon momentum ${\bf P}'={\bf k}$  energy 
\begin{equation}
 E'=k(1-\omega d)\,,
\end{equation}
a null angular momentum, ${\bf L}'=0$, and  ${\bf Q}^{\prime}={\bf k}(1-\omega d)^2$ resulted from Eqs. (\ref{L}) and (\ref{Q}). Then the condition (\ref{disp}) is fulfilled since in this case $m=0$. Furthermore, we derive the conserved quantities measured by the fixed observer $O$ that can be deduced from Eq. (\ref{Kg}) after bringing these quantities in the form (\ref{KK}). After a little calculation we obtain
\begin{eqnarray}
E~&=&k\left[ \sqrt{\frac{1-V}{1+V}}(1-\omega d)-\frac{\omega^2 d^2}{2}\,\frac{V}{\sqrt{1-V^2}}\right]\,,\label{EE}\\
P^1&=&-k\left[ \omega d +\sqrt{\frac{1-V}{1+V}}(1-\omega d)+\frac{\omega^2 d^2}{2}\left(\frac{1}{\sqrt{1-V^2}}-1 \right)\right]\,,\label{PP}\\
Q^1&=&-k\left[-\omega d +\sqrt{\frac{1-V}{1+V}}(1-\omega d)+\frac{\omega^2 d^2}{2}\left(\frac{1}{\sqrt{1-V^2}}+1 \right)\right]\,,~~~~\label{QQ}
\end{eqnarray}
while $P^2=P^3=Q^2=Q^3=0$ and ${\bf L}=0$. These components satisfy the condition $(\ref{Kg})$ for $m=0$.  

The parameters involved in these equations satisfy the natural conditions  $\omega d<1$ and $V<1$ but which are not enough for assuring positive energies observed by $O$.  Therefore, we must impose the supplemental condition $E>0$ which restricts the relative velocity  as 
\begin{equation}\label{VV}
V<V_{lim}(d)=\frac{2(1-\omega d)}{1+(1-\omega d)^2}\,.
\end{equation}
This is in fact the mandatory condition for observing the photon in $O$ at finite time. When the relative velocity $V$ exceeds this limit then the photon cannot arrive in  $O$ at finite time because of the background expansion. Thus $V_{lim}(d)$ defines a new velocity horizon restricting the velocities such that for very far sources, with $\omega d\sim 1$, this limit vanishes.  

\subsection{Redshift and related quantities}

Eq. (\ref{EE}) allows us to derive the formula of the longitudinal Doppler effect in de Sitter relativity. Denoting with $\nu_0$  the frequency of the emitted photon and by $\nu$ that measured by the observer $O$ we can rewrite Eq. (\ref{EE}) as
\begin{equation}\label{nu} 
\frac{1}{1+z}=\frac{\nu}{\nu_0}=\frac{E}{k}= \sqrt{\frac{1-V}{1+V}}\left(1-\omega d -\frac{\omega^2 d^2}{2}\,\frac{V}{1-V}\right)\,,
\end{equation}
pointing out the redshift $z$.  What is new here is the last term proportional with $\omega^2 d^2$ which is generated by the fact that we used the Lorentzian isometries of our de Sitter relativity instead of the usual Lorentz boosts of  special relativity. This term is important since it cannot be seen as a mere small correction as long as for relative velocities approaching to $V_{lim}(d)$ the redshift can increase significantly.  

Obviously, dropping out this new term we recover the well-known factorization  obtained when  the special relativity is used for calculating the Doppler effect.  Note that this supplemental term vanishes in both the particular cases of $V=0$ or $d=0$. Therefore, for $V=0$ we obtain the Lema\^ itre redshift \cite{L1,L2}
\begin{equation}
1+z=\frac{1}{1-\omega d}=\frac{a(t_{c\,obs})}{a(t_{c0})}=\frac{t_{c0}}{t_{c\, obs}}\,,
\end{equation}
since  the photon is emitted at $t_{c0}=t_{c0}'=-\omega^{-1}$ and observed at  $t_{c\, obs}=-{\omega}^{-1}+d$. For $d=0$ we obtain the familiar formula of the Doppler effect in special relativity  since  Eq. (\ref{nu}) depends only on the product $\omega d$ such that the limit $d\to 0$ is equivalent to the flat limit $\omega\to 0$.

In de Sitter relativity the conserved quantities satisfy the condition (\ref{disp}) which can be interpreted as a Lorentz violation produced by the de Sitter gravity. Under such circumstances we expect to find new dispersion relations depending on the parameters $d$ and $V$. Indeed, from Eqs. (\ref{Eq1}) and (\ref{Eq2}) we deduce the linear dispersion relation $E=|{\bf P}| f(d,V)$ where
\begin{equation}
f(d,V)=\frac{1}{V}\frac{(V\omega d-V+1)^2+V^2-1}{(1-\sqrt{1-V^2})\left[1-(\omega d-1)^2\right]-2(V\omega d -V+1)}\,.
\end{equation}
This function is positively defined, $0<f(d,V)\le 1$, for $0\leq d<1$ and $0\leq V<V_{lim}(d)$  vanishing when $V \to V_{lim}(d)$. Moreover, the expansion 
\begin{equation}
f(d,V)=1+\omega d \sqrt{\frac{1+V}{1-V}}+{\cal O}(\omega^2 d^2)\,,
\end{equation}
shows that  this function  is produced exclusively by the de Sitter gravity where the energy depends on the space position. For this reason we find that for $d\to 0$ we recover the result of special relativity $f(0,V)=\lim_{\omega\to 0}f(d,V)=1$. 

Apart from the study of the conserved quantities it is interesting to find which is the propagation time of the photon as well as the real distance between $O$  and $S$ at the time when the redshift is observed by $O$. It is convenient to start this investigation in the charts with conformal coordinates calculating the time $t_{c\,f}'$ when the photon is measured by $O$. Obviously, this is the solution of the equation ${x}^1_{ph}(t_c')=0$ which can be calculated according to Eqs. (\ref{G2}) and (\ref{Eq2}) as
\begin{equation}\label{tf}
t_{c\,f}'=\frac{(V \omega d-V+1)^2+V^2-1}{2\omega V (V\omega d-V+1)}\,.
\end{equation}
This is a conformal time which must remain negative for any values of parameters. It is not difficult to show that $t_{c\,f}'<0$ only if $V$ satisfies the condition  (\ref{VV}) which guarantees that the photon arrives in $O$. Eq. (\ref{tf}) helps us to find the propagation time in the frame with de Sitter-Painlev\' e coordinates by using the function (\ref{Eq1}) as
\begin{equation}
t_f=-\frac{1}{\omega}\ln\left\{-\omega\, t_c[t'_{c\,f},{\bf x}'_{c\,ph}(t'_{c\,f})]\right\}
\end{equation}
since in this frame the photon was emitted at $t_0=0$. This time can be expanded as 
\begin{equation}
t_f=\sqrt{\frac{1+V}{1-V}}\left(d+\frac{\omega d^2}{2} +\frac{\omega^2 d^3}{6}\frac{2-V}{1-V}+{\cal O}(\omega^3)\right)\,,
\end{equation}
pointing out that in the flat limit, when $\omega \to 0$, we obtain just the propagation time predicted in special relativity. 

Now we can derive the distance between $O$ and $S$ at the moment when the photon is observed in $O$. This can be calculated either in the conformal chart $\{t_c,{\bf x}_c\}$ ($D_c$) or in the de Sitter-Painlev\' e one ($D$) as 
\begin{equation}
D_c=x^1_c(t'_{c\,f},d,0,0)~\to~ D=-\frac{D_c}{\omega t'_{c\,f}}\,,
\end{equation}
where the function $x^1_c(t'_{c\,f},d,0,0)$ is given by Eq. (\ref{Eq2}) in which we substitute $t'_c=t'_{c\,f}$ and ${\bf x}'_c=(d,0,0)$. The resulted physical distance $D$ is a complicated function of $(d, v, \omega)$ which cannot be written here. Nevertheless, by using a suitable algebraic code on computer we find the expansion
\begin{eqnarray}
D&=&\frac{d}{(1-V)\sqrt{1-V^2}}\nonumber\\
&+&\omega d^2\, \frac{2V^2-2V-2+\sqrt{1-V^2}(4-3V^2)}{2(1-V^2)(1-V)^2} +{\cal O}(\omega^2)\,,
\end{eqnarray}
where the first term is just the distance calculated in special relativity. Another useful expansion can be obtained for small relative velocities as
\begin{eqnarray}
D&=&\frac{d}{1-\omega d}+\frac{d\,V}{2}\left[1+\frac{1}{(1-\omega d)^2}\right]\nonumber\\
 &+&\frac{d\,V^2}{4} \left[3(1-\omega d)+\frac{2}{1-\omega d}+\frac{1}{(1-\omega d)^3}\right] +{\cal O}(V^3)\,.
\end{eqnarray} 
Here the first term is given by the de Sitter gravity in the absence of the relative motion. These approximations will help one to use simpler analytical formulas in some domains of variables but for results of higher accuracy the numerical calculation cannot be avoided.

\section{Concluding remarks}

We derived the formula of the longitudinal Doppler effect in the de Sitter expanding universe.  The redshift formula obtained with the help of our new Lorentzian isometries \cite{CdSR1} combines the cosmological and kinetic contribution in a non trivial manner in a new term proportional with $\omega d V$. Consequently,  when we eliminate one of these contributions, taking either $V=0$ or $d=0$  this term vanishes allowing us to recover the particular cases of Lema\^ itre's  law or the Doppler effect of special relativity. In addition, we have seen that the related quantities, the dispersion relation, the propagation time and the final distance to the source, have specific features determined by the de Sitter gravity but becoming in the flat limit just those of special relativity.

The principal results reported here, the redshift formula and the dispersion relation, depend only on the conserved quantities which are independent on the local chart we choose. The other related quantities are obtained in conformal coordinates since these are suitable for studying the affects of the space translations as that between $O'$ and $S$   depending on the parameter $d$. Furthermore, for obtaining formulas that may be helpful in interpreting astronomical measurements we have rewrote these results in de Sitter-Painlev\' e coordinates substituting  the conformal coordinates according to Eqs. (\ref{EdS}). This method seems to be flexible and appropriate for studying the effects of the relative motion in de Sitter relativity.   The next challenge is to consider the general  case when the source has an arbitrary position in the mobile frame generating transverse effects.

\end{document}